\documentclass[twocolumn]{elsart3p}
\usepackage{graphicx}
\usepackage{bm}

\journal{Planetary and Space Science}

\begin{document}

\begin{frontmatter}

\title{
Cross-scale coupling at a perpendicular collisionless shock
}

\author{Takayuki Umeda},
\ead{umeda@stelab.nagoya-u.ac.jp}
\author{Masahiro Yamao}
\address{Solar-Terrestrial Environment Laboratory, Nagoya University, 
Nagoya 464-8601, JAPAN}

\author{Ryo Yamazaki}
\address{Department of Physical Science, Hiroshima University, 
Higashi-Hiroshima 739-8526, JAPAN\\
Present address: Department of Physics and Mathematics,
Aoyama Gakuin University, 5--10--1, Fuchinobe, Sagamihara, Kanagawa, 252-5258, JAPAN
}

\begin{abstract}
A full particle simulation study is carried out 
on a perpendicular collisionless shock 
with a relatively low Alfven Mach number ($M_A = 5$). 
Recent self-consistent hybrid and full particle simulations 
have demonstrated ion kinetics are essential 
for the non-stationarity of perpendicular collisionless shocks, 
which means that physical processes due to ion kinetics 
modify the shock jump condition for fluid plasmas. 
This is a cross-scale coupling between 
fluid dynamics and ion kinetics. 
On the other hand, it is not easy to study 
cross-scale coupling of electron kinetics with 
ion kinetics or fluid dynamics, 
because it is a heavy task to conduct 
large-scale full particle simulations of collisionless shocks. 
In the present study, we have performed 
a two-dimensional (2D) electromagnetic full particle simulation 
with a ``shock-rest-frame model''. 
The simulation domain is taken to be larger 
than the ion inertial length 
in order to include full kinetics of both electrons and ions. 
The present simulation result has confirmed the 
transition of shock structures from 
the cyclic self-reformation to the quasi-stationary shock front. 
During the transition, electrons and ions are thermalized 
in the direction parallel to the shock magnetic field.
Ions are thermalized by low-frequency electromagnetic waves 
(or rippled structures) excited by strong ion temperature anisotropy 
at the shock foot, while 
electrons are thermalized by high-frequency electromagnetic waves 
(or whistler mode waves) excited by electron temperature anisotropy 
at the shock overshoot. 
Ion acoustic waves are also excited at the shock overshoot 
where the electron parallel temperature becomes higher than 
the ion parallel temperature. 
We expect that ion acoustic waves are responsible for 
parallel diffusion of both electrons and ions, and that 
a cross-scale coupling between 
an ion-scale mesoscopic instability and 
an electron-scale microscopic instability 
is important for structures and dynamics of 
a collisionless perpendicular shock. 
\end{abstract}

\begin{keyword}
collisionless shock; 
particle-in-cell simulation; 
cross-scale coupling
\end{keyword}

\end{frontmatter}

\newcommand{\Vec}[1]{\mbox{\boldmath $#1$}}
\newcommand{\dpart}[2]{\frac{\partial {#1}}{\partial {#2}}}
\newcommand{\ddpart}[2]{\frac{\partial^2 {#1}}{\partial {#2}^2}}


\section{Introduction}

Dynamics of shock waves in plasmas are often discussed 
by the shock jump conditions (Rankine-Hugoniot conditions), 
which describe conservation laws of mass, momentum, energy, 
normal magnetic field and motional electric field 
for fluid plasmas. 
On the other hand, previous kinetic simulations revealed that 
collisionless shocks in plasmas can be strongly non-stationary
in both spatial and temporal scales of ions. 
%
%
In the direction normal to the shock surface of 
a quasi-perpendicular collisionless shock, 
a new shock front periodically appears 
(e.g., Biskamp and Welter, 1972; Quest, 1985; Lembege and Dawson, 1987; 
Lembege and Savoini, 1992; Hellinger et al., 2002), 
which is called the self-reformation. 
Incoming ions are reflected upstream at the shock ramp of 
a supercritical quasi-perpendicular collisionless shock, 
and they form a foot in front of the ramp during their gyration.
At the upstream edge of the foot, 
ions are accumulated in time and are reflected upstream, 
which are responsible for the self-reformation. 
The cyclic self-reformation is due to ion dynamics, 
although this process has been confirmed 
in both electromagnetic hybrid and full particle simulations.
In addition, recent full particle simulations have shown that 
electron-scale micro instabilities, 
such as Buneman instability (e.g., Shimada and Hoshino, 2000) 
and modified two-stream instability (e.g., Scholer et al., 2003) 
are excited at the foot during the cyclic self-reformation. 
Scholer and Matsukiyo (2004) has also demonstrated that 
the modified two-stream instability is also 
responsible for the self-reformation. 
Another mechanism of the self-reformation is 
steepening of whistler mode waves 
in upstream regions of oblique shocks (Krasnoselskikh et al., 2002).

In the shock-tangential direction, 
on the other hand, 
there appear fluctuations in the spatial scale of 
ion inertial length in the direction parallel to the shock magnetic field 
(Winske and Quest, 1988; Lowe and Burgess, 2003) 
or ion gyro radius of reflected ions 
in the direction perpendicular to the shock magnetic field 
(Burgess and Scholer, 2007), 
which are called the ``ripples''. 
The compression of incoming ions at collisionless shocks 
results in their adiabatic heating in the shock-normal direction. 
In quasi-perpendicular shocks, however, 
the ion heating in the shock-normal direction 
is more dominated by the gyration of reflected ions. 
Thus an ion temperature anisotropy 
between shock-normal and shock-tangential directions 
becomes a common feature in the transition region of 
quasi-perpendicular shocks. 
The dynamic rippled character of the shock surface 
is thought to be related to the ion temperature anisotropy. 
Although this process has been confirmed in two-dimensional (2D) 
electromagnetic hybrid particle simulations, 
it is difficult to take into account 
the dynamic rippled character of the shock surface 
in 2D electromagnetic full particle simulations. 
This is because current computer resources are not necessarily enough 
to take such a large simulation domain of several ion inertial length. 

Very recently, however, there are several attempts of 
2D electromagnetic full particle simulations 
that take into account 
ion dynamics in both shock-normal and shock-surface directions 
(Hellinger et al. 2007; Amano and Hoshino, 2009; Lembege et al., 2009). 
These results indicate that ion-scale fluctuations 
at perpendicular collisionless shocks can dynamically change 
electron-scale processes such as 
wave excitation and electron acceleration. 
The purpose of this paper is to examine 
a cross-scale coupling between 
the dynamic rippled character of the shock surface 
and electron-scale micro instabilities. 
In order to take into account 
ion dynamics in both shock-normal and shock-surface directions, 
a large-scale 2D electromagnetic full particle simulation 
is carried out by using the ``shock-rest-frame model''.

\section{Full Particle Simulations}

\subsection{Shock-Rest-Frame Model}

There are several different methods for exciting 
collisionless shocks in kinetic simulations of plasmas. 
These include the injection method (or the reflection/wall method) 
(e.g., Quest, 1985; Winske and Quest, 1988; 
Shimada and Hoshino, 2000; Hellinger et al. 2002; 
Lowe and Burgess, 2003; Scholer et al., 2003; 
Burgess and Scholer, 2007; Amano and Hoshino, 2009). 
the plasma release method (Ohsawa, 1985),
and the magnetic piston method 
(e.g., Biskamp and Welter, 1972; Lembege and Dawson, 1987; 
Lembege and Savoini, 1992).
In these methods, collisionless shocks are excited by 
an interaction between a supersonic plasma flow and 
a resting plasma. 
The simulation domain is taken in the downstream rest frame 
with the injection method, while 
the simulation domain is taken in the upstream rest frame 
withe the plasma release and magnetic piston methods. 
Thus an excited shock wave propagates upstream 
in these methods. 
There is also another method called the flow-flow method 
for exciting collisionless shocks (e.g., Omidi and Winske, 1992). 
Since collisionless shocks are excited by 
an interaction between two supersonic plasma flows 
in this method,
there exist forward and reverse shock waves. 
A big problem in these methods is that 
excited collisionless shock waves propagate at a fast velocity, 
and it is necessary to take a very long simulation domain 
in the propagation direction of the shock waves 
in order to follow a long-time evolution of the shock waves. 
This makes it difficult to perform multidimensional simulations 
even with current supercomputer systems. 

An alternative is to excite collisionless shocks 
in the shock rest frame with the ``relaxation method'', 
whereby collisionless shocks are excited by 
an interaction between a supersonic plasma flow and 
a subsonic plasma flow moving in the same direction. 
This method was first used in hybrid particle simulations in 1980's 
(e.g., Leroy et al., 1981, 1982), 
and then in full particle simulations in 1990's 
(Pantellini et al., 1992; Krauss-Varban et al., 1995). 
This method was not so popular 
because of several difficulties in numerical techniques, 
and its application to long-term evolution of shock waves 
was not considered. 
In 2000's, however, long-term 1D simulations with the relaxation method 
have been performed by using Darwin particle code 
(Muschietti and Lembege, 2006) 
and full electromagnetic particle code (Umeda and Yamazaki, 2006). 
Very recently, the relaxation method has also been 
applied to long-term 2D full electromagnetic particle simulations 
(Umeda et al., 2008, 2009). 
In general, 
it is not easy to perform a large-scale (ion-scale) multidimensional 
full electromagnetic particle simulations of collisionless shocks 
even with present-day supercomputers. 
Hence the shock-rest-frame model is important 
to be able to follow the evolution of shock waves 
for a long term with a limited computer resource.

\subsection{Simulation Setup}

We use a 2D full electromagnetic particle code 
(Umeda, 2004), in which 
the full set of Maxwell's equations and 
the relativistic equation of motion for individual electrons and ions 
are solved in a self-consistent mannar. 
The continuity equation for charge is also 
solved to compute the exact current density 
given by the motion of charged particles (Umeda et al., 2003). 

The initial state consists of two uniform regions 
separated by a discontinuity. 
In the upstream region that is taken in the left hand side 
of the simulation domain, 
electrons and ions are distributed uniformly in space and 
are given random velocities $(v_x,v_y,v_z)$ to approximate 
shifted Maxwellian momentum distributions 
with the drift velocity $u_{x1}$, 
number density $n_{1} \equiv \epsilon_0 m_e \omega_{pe1}^2 / e^2$, 
isotropic temperatures $T_{e1} \equiv m_e v_{te1}^2$ and 
$T_{i1} \equiv m_i v_{ti1}^2$, 
where $m$, $e$, $\omega_{p}$ and $v_{t}$ are 
the mass, charge, plasma frequency and 
thermal velocity, respectively. 
Subscripts ``1'' and ``2'' denote 
``upstream'' and ``downstream'', respectively.
The upstream magnetic field $B_{y01} \equiv -m_e \omega_{ce1}/e$ 
is also assumed to be uniform, where $\omega_{c}$ 
is the cyclotron frequency (with sign included). 
The downstream region taken in the right-hand side 
of the simulation domain is prepared similarly with 
the drift velocity $u_{x2}$, density $n_{2}$, 
isotropic temperatures $T_{e2}$ and $T_{i2}$, 
and magnetic field $B_{y02}$. 

We take 
the simulation domain in the $x$-$y$ plane 
and assume a perpendicular shock (i.e., $B_{x0}=0$). 
Since the ambient magnetic field is taken in the $y$ direction, 
free motion of particles along the ambient magnetic field 
is taken into account. 
As a motional electric field, a uniform external electric field 
$E_{z0} =-u_{x1}B_{y01} =-u_{x2}B_{y02}$ 
is applied in both upstream and downstream regions, 
so that both electrons and ions drift in the $x$ direction. 
At the left boundary of the simulation domain in the $x$ direction,
we inject plasmas with the same quantities 
as those in the upstream region, 
while plasmas with the same quantities as those 
in the downstream region are also injected from the right boundary 
in the $x$ direction.
We adopted absorbing boundaries 
to suppress non-physical reflection of electromagnetic waves at 
both ends of simulation domain in the $x$ direction (Umeda et al., 2001), 
while the periodic boundaries are imposed 
in the $y$ direction.

In the relaxation method, 
the initial condition is given by solving 
the shock jump conditions (Rankine-Hugoniot conditions) for 
a magnetized two-fluid isotropic plasma
consisting of electrons and ions (Hudson, 1970). 
In order to determine a unique initial downstream state, 
we need given upstream quantities 
$u_{x1}$, $\omega_{pe1}$, $\omega_{ce1}$, $v_{te1}$, and 
$v_{ti1}$ 
and an additional parameter. 
We assume a low-beta and weakly-magnetized plasma 
such that $\beta_{e1}=\beta_{i1}=0.125$ and 
$\omega_{ce1}/\omega_{pe1}=-0.1$ in the upstream region. 
We also use a reduced ion-to-electron mass ratio $m_i/m_e = 25$ 
for computational efficiency. 
The light speed $c/v_{te1}=40.0$ and 
the bulk flow velocity of the upstream plasma $u_{x1}/v_{te1}=4.0$ 
are also assumed.
Then, the Alfv\'{e}n Mach number is calculated as
$M_A = (u_{x1}/c)|\omega_{pe1}/\omega_{ce1}|\sqrt{m_i/m_e}=5.0$. 
The ion-to-electron temperature ratio 
in the upstream region is given as $T_{i1}/T_{e1}=1.0$. 
In this study, downstream ion-to-electron temperature ratio 
$T_{i2}/T_{e2} = 8.0$ is also assumed as 
another initial parameter 
to obtain the unique downstream quantities 
by solving the shock jump conditions, 
$\omega_{pe2}/\omega_{pe1} = 1.8372$, 
$\omega_{ce2}/\omega_{pe1} = 0.3375$, 
$u_{x2}/v_{te1} = 1.1851$, 
and $v_{te2}/v_{te1} = 2.6393$.

In this study, we perform two runs with different 
sizes of the simulation domain. 
We use $N_x \times N_y = 2048 \times 1024$ 
cells for the upstream region and 
$N_x \times N_y = 2048 \times 1024$ 
cells for the downstream region, respectively, 
in Run A. 
The grid spacing and time step of the present simulation are 
$\Delta x/\lambda_{De1} = 1.0$ and $\omega_{pe1}\Delta t=0.0125$, 
respectively. 
Here $\lambda_{De1}$ is the electron Debye length upstream.
Thus the total size of the simulation domain is 
$10.24l_i \times 5.12l_i$ 
which is long enough to include the ion-scale rippled structure, 
where $l_i = c/\omega_{pi1}(=200\lambda_{De1})$ is the 
ion inertial length. 
In Run B, we use $N_x \times N_y = 2048 \times 128$ 
cells for the upstream region and 
$N_x \times N_y = 2048 \times 128$ 
cells for the downstream region, respectively. 
Thus the the total size of the simulation domain is 
$10.24l_i \times 0.64l_i$, 
in which ion-scale processes along the ambient magnetic field 
is neglected. 
We used 16 pairs of electrons and ions per cell in the upstream region 
and 64 pairs of electrons and ions per cell in the downstream region, 
respectively, at the initial state.

\section{Results}


Figure 1 shows the tangential component of magnetic field $B_y$ 
as a function of position $x$ and time $t$ 
for Runs A and B. 
The position and time are renormalized by 
the ion inertial length $l_i$ and 
the ion cyclotron angular period $1/\omega_{ci1}$, respectively. 
The magnitude is normalized by the 
initial upstream magnetic field $B_{y01}$. 
In Fig.1, 
the tangential magnetic fields $B_y$ are 
averaged over the $y$ direction, 
which means that fluctuations in the shock-tangential direction 
are neglected. 

In the present shock-rest-frame model, 
a shock wave is excited by the relaxation of 
the two plasmas with different quantities. 
Since the initial state is given by the shock jump conditions 
for a ``two-fluid'' plasma consisting of electrons and ions, 
the kinetic effect is excluded in the initial state 
and the excited shock becomes ``almost'' at rest 
in the simulation domain.
In both runs, the shock front appears and disappears on a timescale 
of the downstream ion gyro period, 
which corresponds to the cyclic self-reformation of 
a perpendicular shock. 
The reformation takes place for more than $\omega_{ci1}t = 12$ 
in Run B, 
while the reformation seems to be less significant 
after $\omega_{ci1} t \sim 8$ in Run A.  
The previous 2D full particle simulations have demonstrated 
the transition from the cyclic self-reformation to 
a ``quasi-stationary'' shock front 
(Hellinger et al., 2007; Lembege et al., 2009), 
which is in agreement with Run A. 
However, it should be noted that the self-reformation 
does take place even after $\omega_{ci1} t \sim 8$ in Run A,  
but on a different timescale, 
when we focus on a local tangential magnetic field.

When the length of the simulation domain 
in the shock-tangential direction 
is shorter than the ion inertial length, 
ion-scale fluctuations along the shock surface (ripples) 
do not appear and the profiles of electromagnetic fields 
become almost one-dimensional (Umeda et al., 2008, 2009), 
and there exists apparent cyclic self-reformation 
of the perpendicular shock as seen in Run B.  
The present result suggests that 
the ion-scale fluctuations in the shock-tangential direction 
play an important role in the sequential appearance of 
non-stationary and quasi-stationary shock fronts.

\begin{figure}[t]
\center
\includegraphics[width=0.5\textwidth]{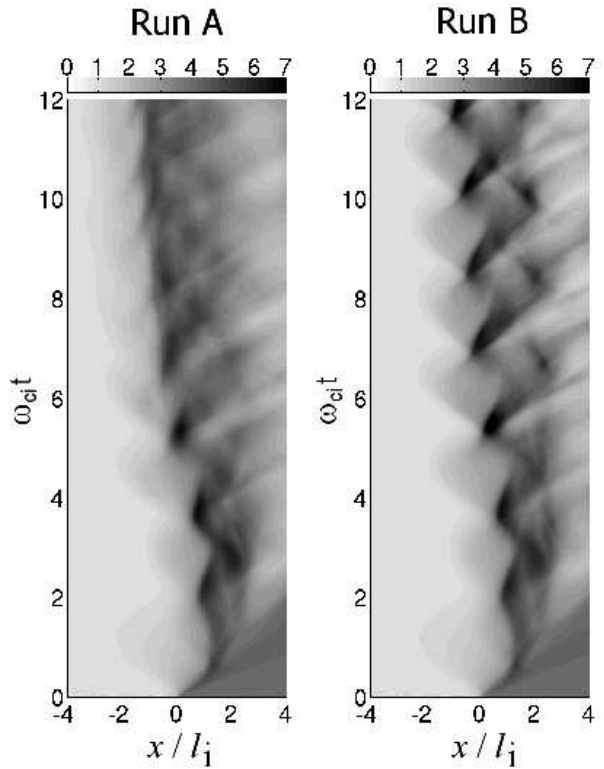}
\caption{
Tangential magnetic field $B_y$ 
as a function of position $x$ and time $t$ 
for Runs A and B. 
The position and time are normalized by 
$\lambda_i$ and $1/\omega_{ci1}$, 
respectively. 
The magnitude is 
normalized by the initial upstream magnetic field $B_{y01}$. 
The magnetic fields are averaged over the $y$ direction. 
}
\end{figure}

\begin{figure*}[t]
\center
\includegraphics[width=1.0\textwidth]{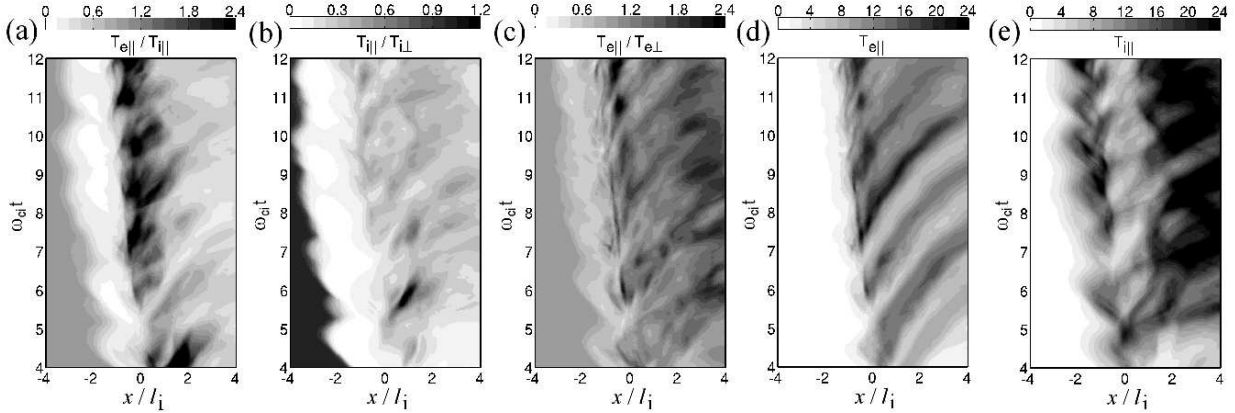}
\caption{
Electron and ion temperatures 
as a function of position $x$ and time $t$ 
for Run A. 
(a)$T_{e||}/T_{i||}$, 
(b)$T_{i||}/T_{i\perp}$, 
(c)$T_{e||}/T_{e\perp}$, 
(d)$T_{e||}$, and (e)$T_{i||}$. 
The position and time are normalized by 
$\lambda_i$ and $1/\omega_{ci1}$, 
respectively. 
The temperatures are normalized by the initial upstream temperature 
($T_{e01}=T_{i01}$). 
These temperatures are averaged over the $y$ direction. 
Here, the parallel temperatures are approximated by 
the temperatures in the $y$ direction while 
the perpendicular temperatures are approximated by the average of 
temperatures in the $x$ and $z$ directions. 
}
\end{figure*}


Figure 2 shows the electron and ion temperatures 
as a function of position $x$ and time $t$ 
for Run A. 
The panels (a), (b) and (c) corresponds to the temperature ratios 
$T_{e||}/T_{i||}$, $T_{i||}/T_{i\perp}$ and $T_{e||}/T_{e\perp}$, 
respectively. 
The panels (d) and (e) corresponds to the parallel temperatures of 
electrons and ions, $T_{e||}$ and $T_{i||}$, respectively. 
Note that these temperatures are averaged over the $y$ direction, 
and that the parallel temperatures are approximated by 
the temperatures in the $y$ direction while 
the perpendicular temperatures are approximated by the average of 
temperatures in the $x$ and $z$ directions.

From $\omega_{ci1} t \sim 7$,
the electron temperature in the direction 
parallel to the ambient magnetic field, $T_{e||}$,
at the shock overshoot 
becomes twice as large as the ion parallel temperature, $T_{i||}$.
At the shock foot, 
the electron perpendicular temperature, $T_{e\perp}$, is 
higher than the electron parallel temperature, $T_{e||}$.
On the other hand, $T_{e||}$ 
becomes higher than $T_{e\perp}$ 
at the shock overshoot.
In the downstream region, $T_{e||}$ 
is slightly higher than 
$T_{e\perp}$.
As seen in Fig.2d,
electrons are strongly thermalized in the direction parallel 
to the shock magnetic field at the overshoot, 
suggesting that 
there exists a strong parallel diffusion process 
at the overshoot from $\omega_{ci1} t \sim 7$. 
On the other hand, ion parallel temperature becomes higher 
at the shock foot and in the downstream region as seen in Fig.2e. 
Fig.2 shows that electrons and ions are thermalized 
in different regions, 
suggesting that electrons and ions heating takes place 
on different scales. 

\begin{figure}[p]
\center
\includegraphics[width=0.5\textwidth]{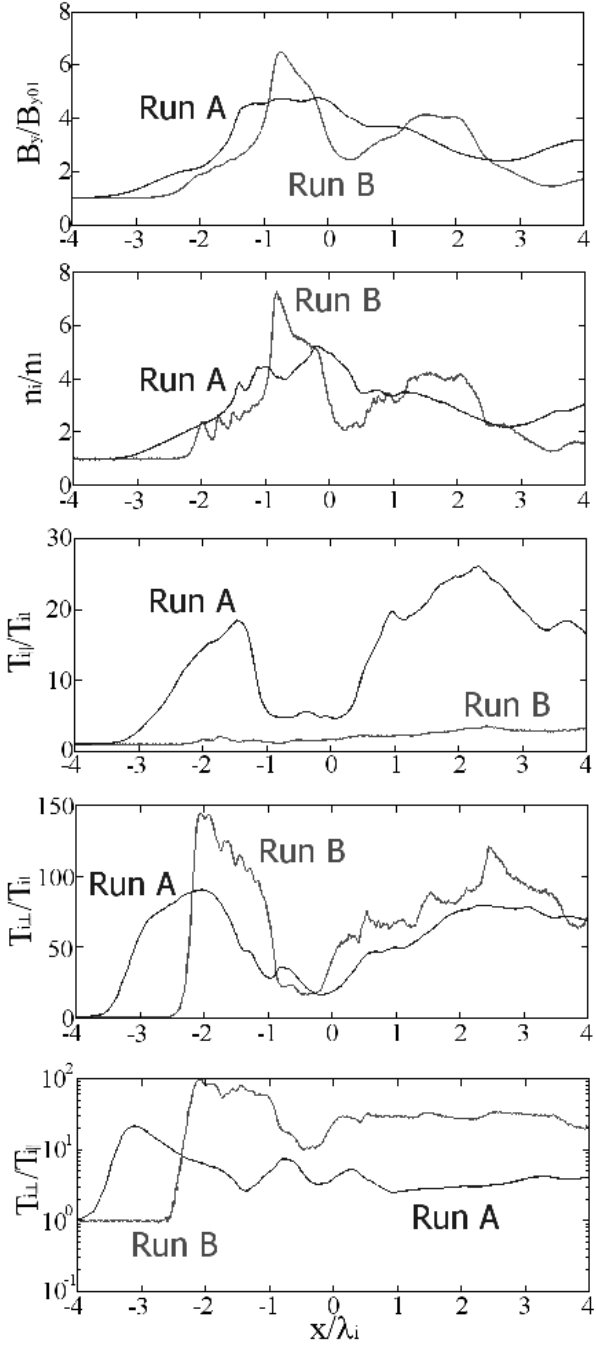}
\caption{
Spatial profiles of 
shock magnetic field $B_y$, ion density $n_i$, 
ion parallel temperature $T_{i||}$, 
ion parallel temperature $T_{i\perp}$, and 
ion temperature ratio $T_{i||}/T_{i\perp}$. 
at $\omega_{ci}t = 12$ for Runs A and B. 
}
\end{figure}

Figure 3 shows snapshots of 
shock magnetic field $B_y$, ion density $n_i$, 
ion parallel temperature $T_{i||}$, 
ion parallel temperature $T_{i\perp}$, and 
ion temperature ratio $T_{i||}/T_{i\perp}$. 
at $\omega_{ci}t = 12$ for Runs A and B. 
Although the Mach number of the present simulation run 
is relatively low ($M_A = 5$), 
the present perpendicular shock is supercritical, and therefore 
the cyclotron motion of reflected ions is dominant for 
ion heating in the shock-normal direction. 
At the shock overshoot, the ion parallel and perpendicular 
temperatures are low because of 
the accumulation of upstream cold ions. 
The ion perpendicular temperature becomes higher 
at the shock foot because of the non-gyrotropic 
velocity distribution of reflected ions. 
At the shock foot, ions are also thermalized in the parallel direction 
because of an anisotropy-driven ion cyclotron wave in Run A. 
However, the ion parallel heating is not responsible for 
the electron parallel heating (see Fig.2). 
As seen in Fig.3, 
the ion temperature anisotropy ($T_{i\perp}/T_{i||}> 1$) would be 
a common feature at perpendicular collisionless shocks. 
In Run B, 
there is no ion parallel heating because 
the system length does not allow the existence of 
an ion cyclotron wave in the parallel direction. 
Thus the temperature anisotropy becomes much higher than in Run A.


In order to study mechanisms for parallel heating of electrons and ions, 
we take Fourier transformation of 
the shock-normal magnetic field component $B_x$ 
and the shock-tangential electric field component $E_y$ 
in the transition region. 
Figure 4 shows frequency-wavenumber spectra of 
the shock-normal magnetic field component $B_x$ 
for different time intervals: 
(a)$\omega_{ci1} t = 4 \sim 6$, 
(b)$\omega_{ci1} t = 6 \sim 8$, 
(c)$\omega_{ci1} t = 8 \sim 10$, and 
(d)$\omega_{ci1} t = 10 \sim 12$. 
The frequency and wavenumber are normalized by 
$\omega_{pe1}$ and $\omega_{pe1}/v_{te1}$, respectively. 
These frequency-wavenumber spectra are obtained by 
projection of $\omega-k_x-k_y$ spectra onto 
the $\omega-k_y$ plane. 
Note that the typical electron and ion cyclotron frequencies 
in the transition region are 
$\omega_{ce} \sim 0.4 \omega_{pe1}$ and 
$\omega_{ci} \sim 0.016 \omega_{pe1}$, respectively, 
and their maximum values are 
$\omega_{ce} \sim 0.7 \omega_{pe1}$ and 
$\omega_{ci} \sim 0.028 \omega_{pe1}$, respectively, 
at the overshoot.

In Fig.4, we found a strong enhancement of $B_x$ component 
below $\omega_{ci}$, which might correspond to 
the rippled structures due to the ion temperature anisotropy. 
We also found an enhancement of $B_x$ component 
over $\omega_{ci}$, suggesting that 
electromagnetic electron cyclotron waves are excited 
in the transition region, 
which might correspond to the ``nonlinear whistler waves'' 
reported by Hellinger et al. (2007) and Lembege et al. (2009). 

For $\omega_{ci1} t = 6 \sim 8$ the high-frequency 
electromagnetic electron cyclotron waves are enhanced 
over $\omega/\omega_{pe1} = 0.4$, while other time intervals 
the $B_x$ component is enhanced up to 
$\omega/\omega_{pe1} = 0.4$, 
implying that the high-frequency waves 
are responsible for the parallel heating of electrons 
at $\omega_{ci}t \sim 7$ (Fig.2a and 2d). 
Note that excitation of electromagnetic whistler mode waves 
due to electron temperature anisotropy 
are observed in the previous 2D simulations of perpendicular shocks 
(Umeda et al., 2008). 
Since the electron parallel temperature becomes 
higher than the electron perpendicular temperature 
for $\omega_{ci}t > 7$, whistler mode waves due to 
electron temperature anisotropy. 

Ions are thermalized in the direction parallel 
to the shock magnetic field by 
the low-frequency electromagnetic waves below $\omega_{ci}$. 
However, these waves are not so much responsible for 
the parallel heating of electrons, 
because there is not significant parallel heating of 
electrons at the shock foot as seen in Fig.2d.

\begin{figure}[p]
\center
\includegraphics[width=0.4\textwidth]{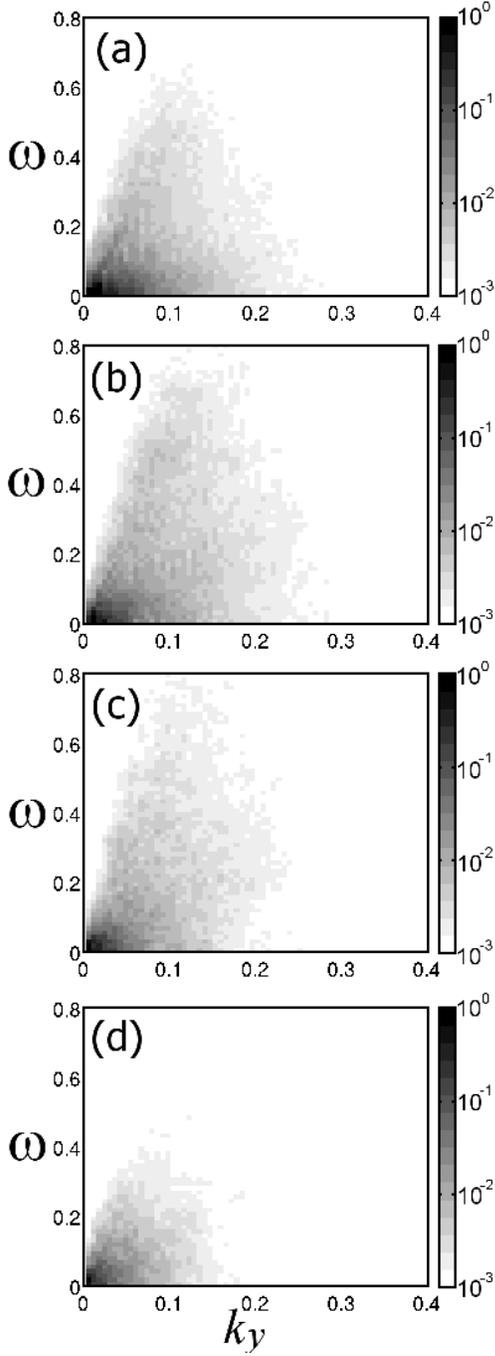}
\caption{
Frequency-wavenumber spectra of 
the shock-normal magnetic field component $B_x$ 
for different time intervals in Run A.  
(a)$\omega_{ci1} t = 4 \sim 6$, 
(b)$\omega_{ci1} t = 6 \sim 8$, 
(c)$\omega_{ci1} t = 8 \sim 10$, and 
(d)$\omega_{ci1} t = 10 \sim 12$. 
The Fourier transformation is taken for 
$y/l_i = 0 \sim 5.12$, 
(a) $x/l_i = -0.5 \sim 0.5$, 
(b)-(d) $x/l_i = -1.0 \sim 0.0$. 
These frequency-wavenumber spectra are obtained by 
projection of $\omega-k_x-k_y$ spectra onto 
the $\omega-k_y$ plain. 
The frequency and wavenumber are normalized by 
$\omega_{pe1}$ and $1/\lambda_{De1}$, respectively. 
The magnitude is normalized by the upstream magnetic field 
$B_{y01}$. 
}
\end{figure}

\begin{figure}[p]
\center
\includegraphics[width=0.4\textwidth]{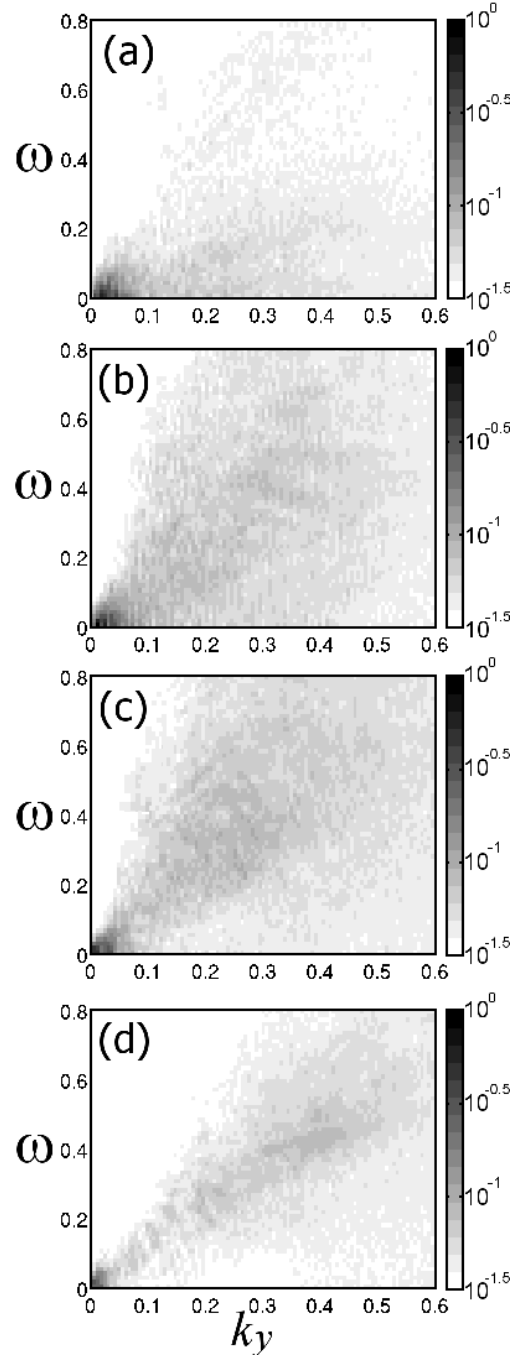}
\caption{
Frequency-wavenumber spectra of 
the shock-normal magnetic field component $E_y$ 
for different time intervals in Run A, 
with the same format as Fig.4. 
(a)$\omega_{ci1} t = 4 \sim 6$, 
(b)$\omega_{ci1} t = 6 \sim 8$, 
(c)$\omega_{ci1} t = 8 \sim 10$, and 
(d)$\omega_{ci1} t = 10 \sim 12$. 
The magnitude is normalized by the motional electric field 
$E_{z0}$. 
}
\end{figure}

Figure 5 shows frequency-wavenumber spectra of 
the shock-normal magnetic field component $E_y$ 
in the transition region for different time intervals: 
(a)$\omega_{ci1} t = 4 \sim 6$, 
(b)$\omega_{ci1} t = 6 \sim 8$, 
(c)$\omega_{ci1} t = 8 \sim 10$, and 
(d)$\omega_{ci1} t = 10 \sim 12$. 
with the same format as Fig.4. 
The typical ion plasma frequency 
in the transition region is 
$\omega_{pi} \sim 0.35 \omega_{pe1}$, 
and its maximum value is 
$\omega_{ce} \sim 0.6 \omega_{pe1}$ 
at the overshoot.

For $\omega_{ci1} t = 6 \sim 12$, 
we found a strong enhancement of the $E_y$ component 
up to $\omega/\omega_{pe1} \sim 0.7$, 
while there is not any enhancement 
in a high-frequency range for $\omega_{ci1} t = 4 \sim 6$. 
The phase velocity of these wave are estimated as 
$v_p/v_{te1} = 1.0 \sim 1.5$. 
From Fig.2, the typical parallel temperatures 
of electrons and ions are estimated as $T_{e||} \sim 18T_{e01}$ 
and $T_{e||} \sim 9T_{i01}$. 
Thus the ion acoustic velocity is obtained as 
\begin{equation}
v_s = \sqrt{\frac{T_{e||}+\gamma T_{i||}}{m_i}} \sim 1.3 v_{te1}
\end{equation}
with $\gamma = 3$, 
suggesting that 
the ion acoustic waves are excited in the transition region. 
As shown in Fig.2a, the electron parallel temperature 
becomes higher than the ion parallel temperature 
due to electron cyclotron waves at the shock overshoot, 
which is a suitable condition for excitation of ion acoustic waves. 

Figs.4 and 5 show 
active wave phenomena in the transition region, 
although the shock front appears to be ``quasi-stationary'' 
when averaged over the $y$ direction. 
The shock front becomes turbulent instead of quasi-stationary. 
We expect that the ion acoustic waves would play a role 
in the transition process from the self-reformation phase 
to the turbulent phase, 
because ion acoustic waves are responsible for 
diffusion of both electrons and ions along a magnetic field.

\section{Summary}

We performed a 2D electromagnetic full particle simulation of 
a low-Mach-number perpendicular collisionless shock. 
The results are itemized below. 

\begin{enumerate}

\item
It has been confirmed that the cyclic self-reformation of 
the shock front becomes less significant as time elapses, 
which is consistent with the previous 2D simulations 
(Hellinger et al., 2007; Lembege et al., 2009). 
The shock front appears to be ``quasi-stationary'' 
by averaging the spatial profiles of electromagnetic fields 
over the shock-tangential direction, 
although electron-scale microscopic and ion-scale mesoscopic 
instabilities are quite dynamic 
and the shock front becomes turbulent. 

\item
During the transition from the cyclic self-reformation 
to the turbulent shock front, 
electrons and ions are thermalized in the direction parallel 
to the shock magnetic field 
in different regions and by different mechanisms. 
The electron parallel temperature is more enhanced than 
the parallel ion temperature at the shock overshoot. 

\item
Low-frequency electromagnetic (ion cyclotron or mirror mode) waves 
are excited at the shock foot, which corresponds to 
the rippled structure in the shock-tangential direction. 
These waves are excited by the strong temperature anisotropy of ions. 
Ions are thermalized in the direction parallel to the shock magnetic field 
by these waves, but electrons are not. 

\item
Electromagnetic electron cyclotron (whistler mode) waves 
are excited at the shock overshoot. 
These waves are excited by the temperature anisotropy of electrons. 
Electrons are thermalized in the direction 
parallel to the shock magnetic field by these waves, but ions are not. 

\item
Strong parallel heating of electrons at the shock overshoot 
results in the suitable condition for excitation of ion acoustic waves. 
The rippled structure might be an energy source 
of the ion acoustic waves. 
However, their detailed excitation mechanism is not yet clear. 
The ion acoustic waves are responsible for 
parallel diffusion of both electrons and ions. 
We expect that the ion acoustic waves have 
a direct implication with the 
transition from the self-reformation phase 
to the turbulent phase. 

\item
A cross-scale coupling between 
an ion-scale mesoscopic instability and 
an electron-scale microscopic instability 
is important for structures and dynamics of 
collisionless perpendicular shocks. 
Hence large-scale full kinetic simulations 
are quite important. 

\end{enumerate}

It is noted that Yuan et al. (2009) have demonstrated the 
cyclic self-reformation in their 2D hybrid simulation of 
a quasi-perpendicular shock with $\theta = 85^{\circ}$, 
which is different from the results in purely perpendicular shocks. 
We are now trying to check whether the reformation is suppressed or not 
at a oblique shock by a large-scale 2D PIC simulation. 
However, this is beyond the scope of the present paper. 

Finally, the effect of mass ratio is discussed. 
In the present simulation parameter 
($M_A = 5, \omega_{pe1}/\omega_{ce1} = 10, \beta = 0.125$), 
electron cyclotron (Bernstein) modes is weakly unstable 
due to current driven instability 
with a reduced mass ratio (e.g., Muschietti and Lembege, 2006). 
When we use the real mass ratio of $m_i/m_e = 1836$, 
the thermal velocity of upstream electrons becomes about 8.6 times 
as large as the case of $m_i/m_e = 25$, 
and the electron cyclotron modes are stabilized. 
In contrast with current driven instability, 
obliquely propagating whistler mode waves becomes 
unstable due to modified two-stream instability 
(Matsukiyo and Scholer, 2003, 2006). 
However, the evolution of modified two-stream instability 
at perpendicular shocks has been studied 
only in a localized uniform model (Matsukiyo and Scholer, 2006). 
The influences of modified two-stream instability 
on the reformation process of perpendicular shocks 
is an outstanding issue to be addressed 
by future 2D PIC simulations of perpendicular collisionless shocks.


\ack
The authors are grateful to S. Matsukiyo and Y. Ohira 
for discussions. 
The computer simulations were carried out 
on Fujitsu HPC2500 at ITC in Nagoya Univ. and 
NEC SX-7 at YITP in Kyoto Univ. 
as a collaborative computational research project at 
STEL in Nagoya Univ. and YITP in Kyoto Univ. 
This work was supported by 
Grant-in-Aid for Scientific Research on Innovative Areas No.21200050 (T. U.), 
Grant-in-Aid for Scientific Research on Priority Areas No.19047004 (R. Y.) 
and Grant-in-Aid for Young Scientists (B) No.21740184 (R. Y.) 
from MEXT of Japan.

\end{document}